\documentclass[journal=apchd5,manuscript=article]{achemso}
\usepackage[utf8]{inputenc}
\usepackage{graphicx}
\usepackage{amsmath}
\usepackage{xcolor}
\usepackage{xr}
\title{Surface Plasmon Polariton Excitation in Time-modulated Media}%

\author{T. V. Raziman}
\affiliation{The Blackett Laboratory, Department of Physics, Imperial College London, London SW7 2AZ, UK}
\alsoaffiliation{Department of Mathematics, Imperial College London, London SW7 2AZ, UK.}
\email{t.raziman@imperial.ac.uk}

\author{Marie Touboul}
\affiliation{POEMS,  CNRS, INRIA, ENSTA, Institut Polytechnique de Paris, 91120 Palaiseau, France}
\alsoaffiliation{Department of Mathematics, UMI 2004 Abraham de Moivre-CNRS, Imperial College London, London SW7 2AZ, UK.}
\author{Riccardo Sapienza}
\affiliation{The Blackett Laboratory, Department of Physics, Imperial College London, London SW7 2AZ, UK}
\author{Richard V. Craster}
\affiliation{Department of Mathematics, Imperial College London, London SW7 2AZ, UK.}
\author{Francisco J. Rodr\'iguez-Fortu\~no}
\email{francisco.rodriguez_fortuno@kcl.ac.uk}
\affiliation{Department of Physics, King's College London, Strand, London WC2R 2LS, UK}

\begin{document}
\maketitle

\begin{abstract}

Surface plasmon polaritons (SPPs) are central to application areas such as sensing, energy harvesting, and nanoscale optics, and are typically excited via spatial structuring -- an approach lacking dynamic control. 
We demonstrate that time modulation of a dispersive Drude-like metal allows the excitation and out-coupling of SPPs, thereby modelling realistic transparent conducting oxides; this establishes a pathway for the active control and extraction of plasmons in experimentally viable time-varying systems.
Using numerical simulations we show that time modulation facilitates both the launching and radiation of surface plasmons with frequencies governed by the dispersion of the bounding media.
Our results also reveal the generation of time-reflected waves at step-like temporal interfaces and the emergence of a magnetostatic mode required for matching boundary conditions at the interface. 

\end{abstract}

\maketitle

\section{Introduction}

Modulating material properties in time enables electromagnetic phenomena that are forbidden in static media~\cite{Galiffi_2022}.
Spatially engineered metasurfaces and metamaterials -- which have enabled applications such as flat optics, beam shaping, optical computing, and cloaking~\cite{Meinzer_2014, Devlin_2017, Zheng_2024, Ni_2015} -- are constrained by fundamental principles, including energy conservation and reciprocity.
Time modulation overcomes these limitations and enables frequency manipulation, non-reciprocal devices, and synthetic super-luminal motion~\cite{Zhou_2020, Guo_2019, Harwood_2025}. The recent discovery of high-contrast all-optical modulation driven by nonlinearities in transparent conducting oxides (TCOs) has enabled a wide range of optical experiments with propagating waves~\cite{Alam_2016, Jaffray_2022}. However, time modulation of near-field waves remains largely unexplored~\cite{Valero_2026}. The wave equation imbues near fields with unique properties at the expense of sub-wavelength confinement: they overcome fundamental limits~\cite{Papadakis_2021, Pendry_2000}, exhibit complex three-dimensional polarisations~\cite{Eismann_2021}, and enhance light-matter interactions at high intensity hotspots~\cite{Nicholls_2017, Challener_2009}. Time modulation of the near field can potentially reveal new applications.
Coupling between surface waves and radiation by time modulation has been studied in infinitesimally thin Drude-like current sheets~\cite{Galiffi_2020},
bulk plasmas~\cite{Bakunov_1996, Bakunov_1998}, and waveguides~\cite{Shlivinski_2025}.

\begin{figure}[t!]
    \includegraphics[width=\textwidth]{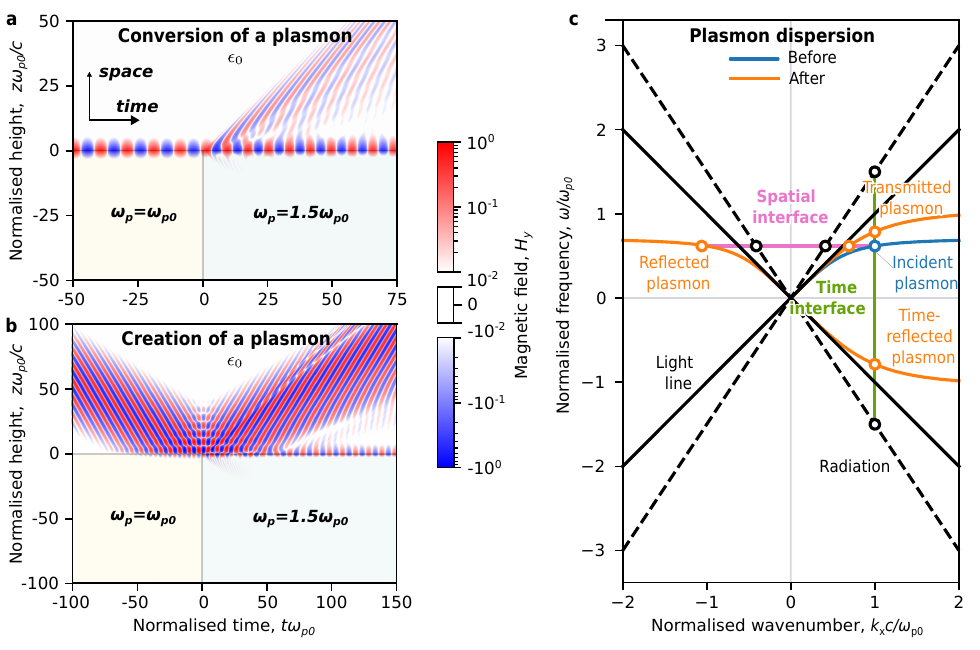}
    \caption{\textbf{Conversion and creation of surface plasmons by time modulation.} (a) A surface plasmon propagates at an interface between a vacuum ($\epsilon=\epsilon_0$, white) and a Drude metal ($\omega_p=\omega_{p0}$, yellow). At $t=0$, the plasma frequency of the metal is changed suddenly to $1.5 \omega_{p0}$ (blue). This changes the frequency of the plasmon and results in plane waves radiating away from the surface. (b) A plane-wave pulse incident from air onto the Drude metal results in the launching of surface plasmons when the plasma frequency is changed abruptly at $t=0$. (c) Mechanism of launching and outcoupling of plasmons. Conventional manipulation of plasmons relies on spatial structuring to compensate for a momentum mismatch (pink). We consider the temporal dual, where a time interface compensates for a frequency mismatch (green).
    }\label{introduction}
\end{figure}

A notable near-field wave is a surface plasmon polariton (SPP), which is a collective excitation of fields and charges bound to a metal-dielectric interface~\cite{Maier_2007}.
SPPs -- and plasmons more generally -- underpin a wide range of technological applications, including chemical and biological sensing, nanoantennas for quantum emitters, solar energy harvesting, on-chip communication, and photocatalysis~\cite{Willets_2007, Kuhn_2006, Catchpole_2008, Kim_2008, Thangamuthu_2022}.
The momentum mismatch of SPPs with propagating waves in the surrounding media [Fig.~\ref{introduction}(c)] has necessitated the development of various coupling schemes~\cite{Maier_2007, Aftab_2024}; however, such spatial schemes are static and limited in their reconfigurability.
Structuration in time gives additional freedom and the space-time duality~\cite{Agrawal_2025} enables the replacement of spatial structuring by time modulation; moreover it lifts the constraint of energy conservation to create new frequencies allowing a propagating wave to couple to a SPP with the same in-plane momentum but now having a different frequency [Fig.~\ref{introduction}(c)]. 

Here, we demonstrate that near-field surface waves can be manipulated and coupled to propagating waves by time modulation of the material properties in realistic platforms of TCOs.
In particular, SPPs can be  excited and out-coupled at the interface of a dispersive Drude-like metal, by a modulation of the carrier mass.
We show that time modulation removes the barrier between propagating waves and the evanescent plasmonic near field, permitting (i) partial out-coupling of surface plasmons into radiated frequencies, and (ii) excitation of a surface plasmon with free-space radiation.
Under step-like modulation, both processes also generate time-reflected~\cite{Mendonca_2002} plasmons and propagating waves, along with a magnetostatic mode in the metal to satisfy boundary conditions.
Our results act to motivate experiments and applications by proposing a mechanism to create, manipulate, and extract plasmons in realistic time-modulated systems.

\section{Results and discussion}

The effect of time modulation on an existing plasmon bound to a metal-dielectric interface is shown in Fig~\ref{introduction}(a) by plotting the real magnetic field $H_y$ as a function of space and time (See Supporting Sections \ref{SI:theory} for the theoretical model and \ref{SI:simulations} for the simulation details). The plasmon lies at the interface between a vacuum ($\epsilon=\epsilon_0$, white) and a Drude metal with background permittivity $\epsilon_\infty=1$, plasma frequency $\omega_{p0}$, and loss $\gamma = 0.02 \omega_{p0}$ (yellow), with fixed momentum along $x$ (here we chose $k_x = \omega_{p0} / c$). Abruptly increasing the plasma frequency of the metal to 1.5$\omega_{p0}$ (blue) at time $t=0$ modifies the fields and we observe: 
the plasmon out-couples partially into radiation after the modulation; the radiation occurs at a range of angles as evident from the phase of the field leaving the interface; 
after the propagating wave has left, part of the plasmon remains bound to the interface with an increased frequency.

Conversion of the plasmon into radiation suggests that the reciprocal effect of creating a plasmon from propagating light should also be possible. In Fig~\ref{introduction}(b) we show this by illuminating a plane wave pulse with fixed momentum ($k_x=0.7\omega_{p0} / c$) on the same system.
The peak of the plane wave pulse falls on the interface at $t=0$, when the plasma frequency is increased:
after the reflected pulse leaves the modulated surface, a plasmon is indeed found at the interface; 
the reflected pulse is also modified from the original, with clear gaps showing the signature of time modulation.

\begin{figure}[t!]
    \includegraphics[width=\textwidth]{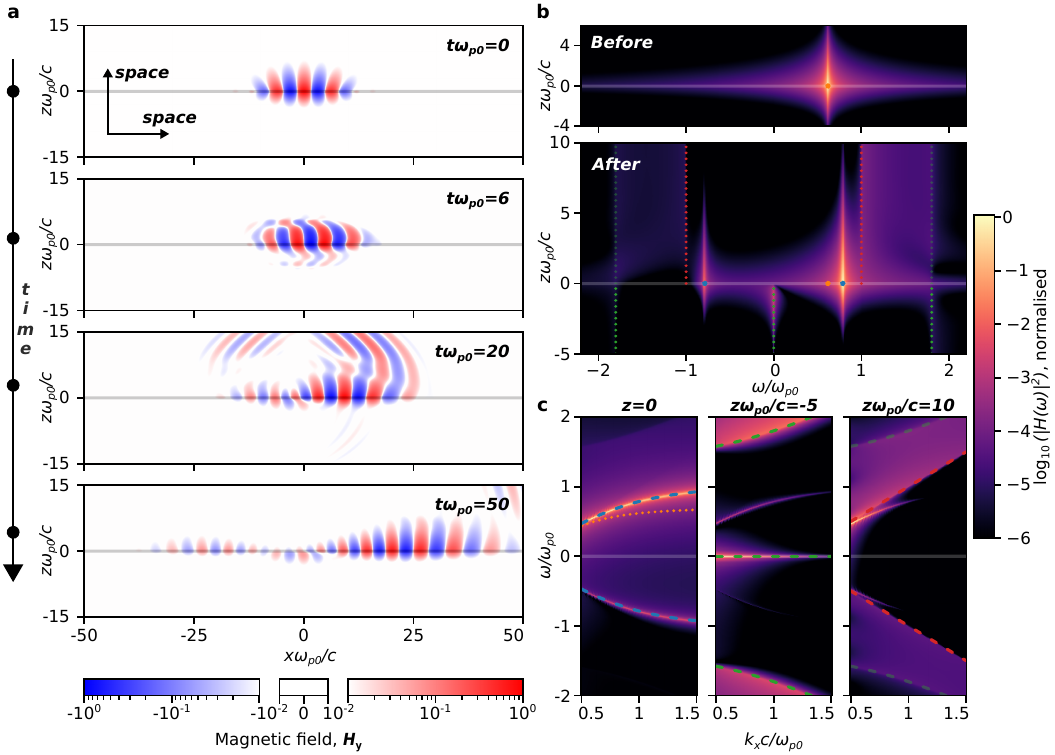}
    \caption{\textbf{Conversion of a surface plasmon pulse by time modulation.} (a) Snapshots of the evolution of the plasmon, starting from modulation at $t=0$. (b) Normalised intensity of frequency spectrum of the initial and converted magnetic field at different heights. Frequency spectrum is evaluated using a truncated Fourier transform. (c) Dispersion of the converted magnetic field as a function of wavevector of the original plasmon.\label{plasmondecay}}
\end{figure}

Figure~\ref{plasmondecay}(a) illustrates the conversion of a realistic plasmon with a finite pulse width in more detail (Animation in Supporting Video SV1).
When the plasma frequency of the metal is changed to $1.5 \omega_{p0}$ at time $t=0$, the plasmon stops being a non-reflecting mode, and starts radiating into both half-spaces.
The field radiated into air leaves as pulses in forward and backward directions, while the field into the metal decays quickly due to losses.
The field that remains bound to the interface splits into three: a part propagating forward, another propagating backwards (time-reflection~\cite{Mendonca_2002}), and the part fixed near the origin (this `frozen mode', a continuous current and its associated magnetostatic field, will be discussed later).
The animation shows that the bound modes have a frequency higher than that of the original plasmon.

The forward and backward propagating modes bound to the interface are surface plasmons as shown in Fig~\ref{plasmondecay}(b); we compute the frequency spectrum  (See Supporting section \ref{SI:spectra} for details) of the magnetic field of the fixed-momentum plasmon of Fig.~\ref{introduction}(a).
The frequency distribution before the modulation is Lorentzian, centered at the analytical SPP frequency (orange dot) and exponentially decaying away from the interface.
The most prominent feature after the modulation is the bound wave that retains its shape and shifts to a higher frequency corresponding to the SPP in the modulated medium (blue dot).
The bound mode that propagates backwards has the same frequency with a negative sign, confirming that it is a time-reflected plasmon.
The frequencies of the radiating field in air range from the light line corresponding to the momentum of the plasmon (red dotted lines) to the dispersion of the bulk metal (green dotted lines).
This observation confirms that the out-coupling of radiation is a consequence of the original plasmon ceasing to be a non-reflecting mode after modulation.
The field distribution of the old plasmon, which had a single real in-plane momentum $k_x$ and an imaginary $k_z$, and whose spatial dependence is preserved immediately after the sudden change in plasma frequency, no longer corresponds to a monochromatic eigenmode solution at the interface of air with the new metal plasma frequency.
Instead, the evanescent spatial distribution is decomposed into propagating and evanescent solutions, all with the same $k_x$ (which is a conserved quantity in this problem due to symmetry under spatial translation~\cite{Liberal_2024, Zhang_2024}) but with a range of frequencies, and hence different values of $k_z$.
Some of these out-of-plane momenta are real, corresponding to propagating waves at different angles, and are thus able to transmit to air as radiation.
These conclusions are further established by plotting the intensities at, above and below the interface for various values of initial plasmon momentum $k_x$ [Figure~\ref{plasmondecay}(c)].
The bound modes and the radiation frequency cutoffs clearly follow the respective dispersion branches.

\begin{figure}[t!]
    \includegraphics[width=\textwidth]{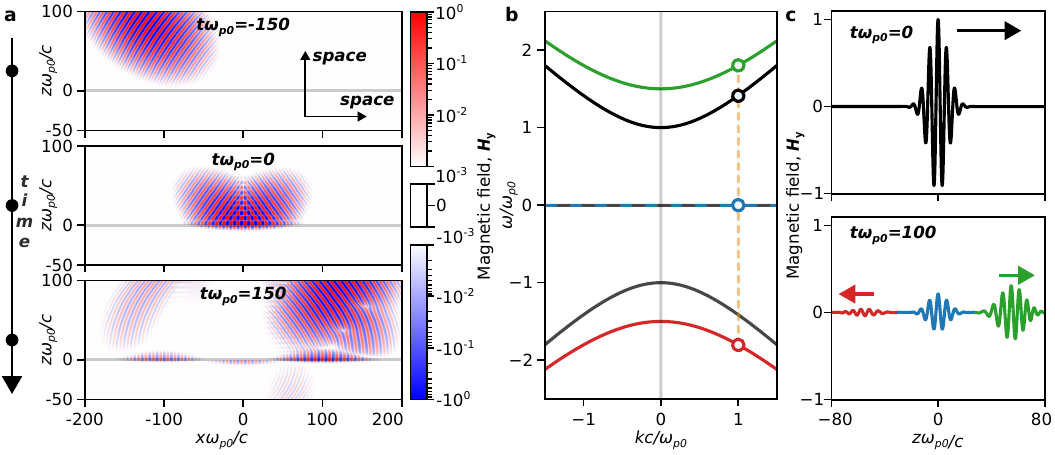}
    \caption{\textbf{Creation of a surface plasmon by time modulation.} (a) Snapshots of a Gaussian pulse incident from air to the metal at $t=0$, when the plasma frequency is changed. (b) Bulk dispersion curves of the medium. When a medium with a forward-propagating wave (black) undergoes time modulation, it decays into three different branches of the medium: forward-propagating (green), time-reflected (red), and frozen (blue). (c) Magnetic field of the pulse that splits into three.\label{plasmoncreation}}
\end{figure}

Next, we study a propagating Gaussian pulse reflecting from a metal surface that is time-modulated at time $t=0$ [Figure~\ref{plasmoncreation}(a). Animation in Supporting Video SV2]. This pulse contains a spectrum of frequencies and momentum components due to the temporal and spatial confinement, and is a realistic model of finite laser pulses in experiments.
Until the modulation, the pulse was getting reflected as if from a metallic mirror.
But after the modulation, the reflected pulse is clearly modified.
There is also a time-reflected pulse that is reflected in the direction of the incident pulse.
After the pulses leave, three bound modes remain on the surface: A forward-propagating plasmon, a time-reflected plasmon, and field that stays fixed to the origin.

We can understand the frozen mode at the origin by studying the bulk dispersion of the metal [Figure~\ref{plasmoncreation}(b)].
Before modulation, the real part of the dispersion of the metal consists of three branches (black lines): two hyperbolas corresponding to forward and backward propagating modes, and a zero frequency mode that is magnetostatic.
When the metal is modulated, hyperbolic branches shift (green and red) but the magnetostatic branch (blue) remains.
To satisfy boundary conditions of electric and magnetic fields as well as the carrier velocity, the original mode needs to decay into all three of these modes.
The zero frequency mode is similar to the magnetostatic field and current that are trapped by a plasma from a propagating field~\cite{Bakunov_1998}, and has been termed the ``wiggler mode'' in time-varying media~\cite{Kalluri_2018, Galiffi_2025}.
This phenomenon is visible in the evolution of a one-dimensional pulse that travels through a bulk metal [Figure~\ref{plasmoncreation}(c)].
After an abrupt modulation, in addition to the forward and time-reflected waves that propagate in the metal with modified frequency, there is also a wave that stays frozen at the origin.
It is the magnetostatic mode that stays at the location of the original pulse while the other branches combine into the new plasmon and the propagating radiation. 

Even when we replace the step-like modulation of $\omega_p$ with a weaker and smoother modulation that has been experimentally achieved in a realistic TCO such as indium tin oxide~\cite{Tirole_2022, Harwood_2025} (see Supporting section~\ref{SI:ITO} for details), surface plasmon pulses can still be converted to radiation (animation in Supporting Video SV3) and excited from propagating waves (animation in Supporting Video SV4).
However, the slowness of modulation restricts the magnitude of frequency shift it can provide to match the various dispersion curves in Fig.~\ref{introduction}(c), suppressing the time-reflection and the frozen mode.
Media with faster modulation comparable to the plasma frequency would be required to observe these effects in realistic systems.

In conclusion, we have demonstrated that time modulation can launch surface plasmons and convert them into propagating waves.
The spectral analysis confirms that the excited bound modes correspond to surface plasmon polaritons and the magnetostatic mode of the medium.
The ability to create and manipulate SPPs on realistic transparent conducting oxides opens the door to practical experiments and devices based on time-varying media.
Although lower frequencies such as the microwave and terahertz regimes do not support SPPs since metals behave as perfect electric conductors, subwavelength structuring can permit ``spoof'' surface plasmons through an effective Drude-like permittivity~\cite{Pendry_2004}.
Similar spoof surface waves exist in acoustoelastics in arrays of perfectly rigid bodies~\cite{Christensen_2014}, and are influenced by space-time modulations~\cite{Pham_2023}.
The mechanism studied here could also be leveraged to launch and convert such spoof surface waves across wave domains.
Further control may be achieved through periodic temporal or travelling-wave modulation that couples spatial and temporal changes, enabling finer manipulation of plasmon dynamics and improving the efficiency.

\begin{suppinfo}
\textit{Supporting information}: Theoretical model; Numerical simulations; Computation of spectra; Modelling realistic indium tin oxide.
\textit{Supporting videos}: Conversion of a plasmon with a finite pulse width (SV1), Excitation of a plasmon from a propagating Gaussian pulse (SV2), Conversion of a plasmon from realistic indium tin oxide (SV3), Excitation of a plasmon from a propagating Gaussian pulse in realistic indium tin oxide (SV4).
\end{suppinfo}

\begin{acknowledgement}
We acknowledge computational resources and support provided by the Imperial College Research Computing Service (http://doi.org/10.14469/hpc/2232).
We acknowledge support from the Engineering and Physical Sciences Research Council (EPSRC), grant number EP/Y015673.
\end{acknowledgement}

\section*{Data availability statement}
The code underlying this study and the parameters to generate the figures are openly available on GitHub.~\cite{Raziman_2026}.

\bibliography{references}

\end{document}